\documentclass[preprint,prc,aps,showpacs,showkeys,groupedaddress,floatfix]{revtex4}
\usepackage{epsfig}
\usepackage{dcolumn}
\usepackage{bm}
\usepackage{graphics}
\usepackage{graphicx}
\usepackage{amssymb}
\usepackage{amsmath}
\begin{document}
\title{Solution of the Radial Schr\"{o}dinger Equation for the Potential Family
$V(r)=\frac{A}{r^{2}}-\frac{B}{r}+Cr^{\kappa}$  using the Asymptotic Iteration Method}
\author{M. Aygun\dag, O. Bayrak\ddag \S \quad and I. Boztosun\S}
\affiliation{\dag Faculty of Arts and Sciences, Department of
Physics, Ataturk University, Erzurum, Turkey}\affiliation{\ddag
Faculty of Arts and Sciences, Department of Physics, Bozok
University, Yozgat, Turkey} \affiliation {\S Faculty of Arts and
Sciences, Department of Physics, Erciyes University, Kayseri,
Turkey}
\date{\today}
\begin{abstract}
We present the exact and iterative solutions of the radial Schr\"
{o}dinger equation for a class of potential,
$V(r)=\frac{A}{r^{2}}-\frac{B}{r}+Cr^{\kappa}$, for various values
of $\kappa$ from -2 to 2, for any $n$ and $l$ quantum states by
applying the asymptotic iteration method. The global analysis of
this potential family by using the asymptotic iteration method
results in exact analytical solutions for the values of $\kappa=0
,-1$ and $-2$. Nevertheless, there are no analytical solutions for
the cases $\kappa=1$ and $2$. Therefore, the energy eigenvalues are
obtained numerically. Our results are in excellent agreement with
the previous works.
\end{abstract}
\keywords{asymptotic iteration method, eigenvalues and
eigenfunctions, Kratzer, Modified Kratzer, Goldman-Krivchenkov,
spiked harmonic oscillator, Coulomb plus linear and Coulomb plus
harmonic oscillator potentials.}\pacs{03.65.Ge} \maketitle
\section{\bf Introduction}
We search for the exact solution of a new class of potential as in
the following form:
\begin{equation}\label{pot}
    V(r) =\frac{A}{r^{2}}-\frac{B}{r}+Cr^{\kappa}
\end{equation}
This potential includes various potentials according to the values
of the potential parameters:

a) By choosing $C=0$, this potential turns into the Kratzer
potential, which is analytically solvable by using various methods
\cite{kratzer,fluge,swkb}. This potential has been used
extensively so far in order to describe the molecular structure and
interactions \cite{kratzer,fluge,swkb}.

b) The Gol'dman-Krivchenkov potential is obtained by choosing $A
\neq 0$, $B=0$ and $\kappa=2$ and it is solved analytically by using
various methods \cite{fluge,envolepe}. Furthermore, this potential
becomes the spiked harmonic oscillator potential \cite{envolepe}
when $A=0$, $B=1$ and $\kappa=2$.

c) The Coulomb plus linear potential form is obtained by choosing
$A=0$ and $\kappa=1$ and is solved by methods such as the envelope
and variational methods \cite{envolepe}. This potential has raised
great interest in atomic and molecular physics (Ref. \cite{Enrique}
and references therein).

d) The Coulomb plus harmonic oscillator potential form is obtained
by choosing $A=0$ and $\kappa=2$. This form is solved by using the
moment method \cite{moment}, the shifted $1/N$ expansion method
\cite{shifted} and the envelope method \cite{envolepe}. In addition,
it is applied to examine the Zeeman quadratic effect \cite{Avron}
and the magnetic field effect in the hydrogen atom \cite{Taut}.

These potentials have been extensively used to describe the bound
and the continuum states of the interactions systems and a great
number of papers have been published for the exact and numerical
solutions of these potentials (see Refs.
\cite{kratzer,fluge,swkb,k1,k2,k3,hakan,envolepe,Enrique,moment,shifted,Avron,Taut}
 and references therein). Thus, it would be interesting and important to solve the
non-relativistic radial Schr\"{o}dinger equation for a new solvable
potential family for any $n$ and $l$ quantum states. Recently, an
alternative method, called as the Asymptotic Iteration Method (AIM),
has been developed by \c{C}ift\c{c}i \emph{et al.} \cite{hakan} for
solving the second-order homogeneous linear differential equations
and it has been applied to solve the non-relativistic radial
Schr\"{o}dinger equation as well as the relativistic wave equations
\cite{hakan,barakat,fernandez,orhan,hakanper,orhanIJQC}. In this
paper, we aim to show that AIM could give the energy eigenvalues for
the potentials that have analytical solutions obtained by using
different methods. Moreover, AIM could provide energy eigenvalues
for potentials that have no analytical solutions for any $n$ and $l$
quantum states with various $\kappa$ values.

In the next section, we introduce AIM and then in section
\ref{analytic}, the analytical solution of the Schr\"{o}dinger
equation is obtained by using AIM for the potentials with $\kappa=0,
-1$ and $-2$ and a closed form for the energy eigenvalues and
corresponding eigenfunctions for any $n$ and $l$ quantum numbers are
given. Then, in section \ref{iterative}, the energy eigenvalues for
a new solvable potential with $\kappa=1$ and $2$ cases by using AIM
iteration procedure are obtained. Finally, in section
\ref{conclude}, we remark on these results.

\section{Overview of the Asymptotic Iteration Method (AIM)}
AIM is proposed to solve the second-order differential equations of
the form \cite{hakan,orhan}.
\begin{equation}\label{diff}
  y''=\lambda_{0}(x)y'+s_{0}(x)y
\end{equation}
where $\lambda_{0}(x)\neq 0$. The variables, $s_{0}(x)$ and
$\lambda_{0}(x)$, are sufficiently differentiable. The differential
equation (\ref{diff}) has a general solution \cite{hakan}
\begin{equation}\label{generalsolution}
  y(x)=exp \left( - \int^{x} \alpha(x^{'}) dx^{'}\right ) \left [C_{2}+C_{1}
  \int^{x}exp  \left( \int^{x^{'}} [\lambda_{0}(x^{''})+2\alpha(x^{''})] dx^{''} \right ) dx^{'} \right
  ]
\end{equation}
if $k>0$, for sufficiently large $k$, we obtain the $\alpha(x)$
values from
\begin{equation}\label{quantization}
\frac{s_{k}(x)}{\lambda_{k}(x)}=\frac{s_{k-1}(x)}{\lambda_{k-1}(x)}=\alpha(x),
\quad k=1,2,3,\ldots
\end{equation}
where
\begin{eqnarray}\label{iter}
  \lambda_{k}(x) & = &
  \lambda_{k-1}'(x)+s_{k-1}(x)+\lambda_{0}(x)\lambda_{k-1}(x) \quad
  \nonumber \\
s_{k}(x) & = & s_{k-1}'(x)+s_{0}(x)\lambda_{k-1}(x), \quad \quad
\quad \quad k=1,2,3,\ldots
\end{eqnarray}

It should be noted that one can also start the recurrence relations
from $k=0$ with the initial conditions $\lambda_{-1}=1$ and
$s_{-1}=0$ \cite{fernandez}. For a given potential, the radial
Schr\"{o}dinger equation is converted to the form of equation
(\ref{diff}). Then, s$_{0}(x)$ and $\lambda_{0}(x)$ are determined
and s$_{k}(x)$ and $\lambda_{k}(x)$ parameters are calculated by the
recurrence relations given by equation~(\ref{iter}).

The energy eigenvalues are obtained from the roots of the
quantization condition, given by the termination condition of the
method in equation (\ref{quantization}). The quantization condition
of the method together with equation (\ref{iter}) can also be
written as follows
\begin{equation}\label{kuantization}
  \delta_{k}(x)=\lambda_{k}(x)s_{k-1}(x)-\lambda_{k-1}(x)s_{k}(x)=0 \quad \quad
k=1,2,3,\ldots
\end{equation}
The energy eigenvalues are obtained from this equation if the
problem is exactly solvable. If not, for a specific $n$ principal
quantum number, we choose a suitable $x_0$ point, determined
generally as the maximum value of the asymptotic wave function or
the minimum value of the potential, and the approximate energy
eigenvalues are obtained from the roots of this equation for
sufficiently great values of $k$ with iteration.

The wave functions are determined by using the following wave
function generator
\begin{equation}
y_{n}(x)=C_2 exp(-\int^{x}\frac{s_{k}(x^{\prime})}{\lambda
_{k}(x^{\prime})}dx^{\prime }) \label{generator1}
\end{equation}
where $k\geq n$, $n$ represents the radial quantum number and $k$
shows the iteration number. For exactly solvable potentials, the
radial quantum number $n$ is equal to the iteration number $k$ and
the eigenfunctions are obtained directly from equation
(\ref{generator1}). For nontrivial potentials that have no exact
solutions,  $k$ is always greater than $n$ in these numerical
solutions and the approximate energy eigenvalues are obtained from
the roots of equation (\ref{kuantization}) for sufficiently great
values of $k$ by iteration.

\section{$\kappa=0, -1$ and $-2$ Cases: Analytical Solutions}
\label{analytic}
Inserting the potential given by equation (\ref{pot}) into the
Schr\"{o}dinger equation gives
\begin{equation}\label{schrana}
\frac{d^{2}R_{nl}}{dr^{2}}+\frac{2m}{\hbar^{2}}\left[E-\frac{A}{r^{2}}+\frac{B}{r}-Cr^{\kappa}-\frac{l(l+1)\hbar^{2}}{2mr^{2}}\right]R_{nl}=0
\end{equation}
where $n$ and $l$ are radial and orbital angular momentum quantum
numbers, $A$, $B$ and $C$ are strictly positive constants. By using the following \emph{ansatze:}
\begin{equation}\label{ansatz}
-\varepsilon^{2} = \frac{2mE}{\hbar^{2}}, \quad
-\epsilon_{nl}^{2}=-\varepsilon_{nl}^{2}-\widetilde{C}, \quad
\widetilde A =\frac{2mA}{\hbar^{2}}, \quad \widetilde B
=\frac{2mB}{\hbar^{2}}, \quad \widetilde C = \frac{2mC}{\hbar^{2}},
\end{equation}
Equation (\ref{schrana}) becomes
\begin{equation}\label{schranaeq}
\frac{d^{2}R_{nl}}{dr^{2}}+\left[-\varepsilon^{2}-\frac{\widetilde{A}}{r^{2}}+\frac{\widetilde{B}}{r}-\widetilde
Cr^{\kappa}-\frac{l(l+1)}{r^{2}}\right]R_{nl}=0
\end{equation}

The aim of this paper is to show how to obtain the analytical and
numerical solutions of
$V(r)=\frac{A}{r^{2}}-\frac{B}{r}+Cr^{\kappa}$ potential with
different $\kappa$ values. $\kappa$=0, -1 and -2 values of this
potential have analytical solutions and have been extensively
studied so far \cite{fluge,kratzer,swkb,orhanIJQC}. Therefore, we do
not give the details of the calculations and only show the
analytical solutions. By inserting $\kappa=0, -1$ and $-2$ into
equation (\ref{schranaeq}) and by doing some simple algebra
described in the previous section, the energy eigenvalues and
regular eigenfunctions are obtained as follows:

\begin{flushleft}
\begin{small}
\begin{tabular}{lll} \hline \hline
$\kappa$ & Eigenvalues  & Eigenfunctions
\\\hline
0, &
$E_{nl}=C-\frac{mB^{2}}{2\hbar^{2}}\left(n+\frac{1}{2}+\sqrt{(l+\frac{1}{2})^{2}+\frac{2mA}{\hbar^{2}}}\right)^{-2}$,
& $R_{nl}(r)=Ne^{-\epsilon_{nl}^{\kappa=0} r}\
_{1}F_{1}(-n,2\Lambda^{\kappa=0}
+2;2\epsilon_{nl}^{\kappa=0} r)$  \\
-1, &
$E_{nl}=-\frac{m(B-C)^{2}}{2\hbar^{2}}\left(n+\frac{1}{2}+\sqrt{(l+\frac{1}{2})^{2}+\frac{2mA}{\hbar^{2}}}\right)^{-2}$,
& $R_{nl}(r)=N e^{-\varepsilon_{nl}^{\kappa=-1}
r}\ _{1}F_{1}(-n,2\Lambda^{\kappa=-1}+2;2\varepsilon_{nl}^{\kappa=-1}r)$ \\
-2, &
$E_{nl}=-\frac{mB^{2}}{2\hbar^{2}}\left(n+\frac{1}{2}+\sqrt{(l+\frac{1}{2})^{2}+\frac{2m}{\hbar^{2}}(A+C)}\right)^{-2}$,
& $R_{nl}(r)=N e^{-\varepsilon_{nl}^{\kappa=-2}
r}\ _{1}F_{1}(-n,2\Lambda^{\kappa=-2}+2;2\varepsilon_{nl}^{\kappa=-2}r)$ \\
\hline\hline
\end{tabular}
\end{small}
\end{flushleft}

with
\begin{equation}
\begin{array}{ll}
\epsilon_{nl}^{\kappa=0}=\frac{\widetilde{B}}{2(n+\Lambda+1)}, & \Lambda^{\kappa=0}=-\frac{1}{2}+\sqrt{(l+\frac{1}{2})^{2}+\widetilde A} \\
\varepsilon_{nl}^{\kappa=-1}=\frac{\widetilde{B}-\widetilde{C}}{2(n+\Lambda+1)},
&
\Lambda^{\kappa=-1}=-\frac{1}{2}+\sqrt{(l+\frac{1}{2})^{2}+\widetilde
A}
\\\varepsilon_{nl}^{\kappa=-2}=\frac{\widetilde{B}}{2(n+\Lambda+1)}, & \Lambda^{\kappa=-2}=-\frac{1}{2}+\sqrt{(l+\frac{1}{2})^{2}+ \widetilde{A}+\widetilde{C}} \\
\end{array}
\end{equation}
These results are in excellent agreement with the previous results
obtained by using different methods (see
\cite{fluge,kratzer,swkb,orhanIJQC} and  references therein).

\section{$\kappa=1$ and $2$ Cases: Iterative Solutions}
\label{iterative} For cases $\kappa=1$ and $2$, the exact analytical
solutions can not be found and in this section, we present how to
find the energy eigenvalues by applying the asymptotic iteration
method. If we consider $V(r)=\frac{A}{r^2}-\frac{B}{r}+Cr^{\kappa}$
potential in the three-dimensional radial Schr\"{o}dinger equation,
we obtain equation (\ref{schrana}). The straightforward application
of AIM to solve this equation gives us the energy eigenvalues,
however, we have observed that the energy eigenvalues oscillate and
do not converge within a reasonable number of iteration. The
sequence appears to converge when the number of iterations $k \simeq
30$, but then it begins to oscillate as the iteration number $k$
increases. This result violates the principle behind the AIM; as the
number of iteration increases, the method should converge and should
not oscillate. In order to overcome this problem and obtain a rapid
convergence, we make a change of variables as $r=r_{0}\rho$, where
$r_{0}=\frac{\hbar^{2}}{2mB}$, then we obtain
\begin{equation}\label{schr}
\frac{d^{2}R_{nl}}{d\rho^{2}}+
\left[\varepsilon+\frac{1}{\rho}-\gamma^{2}\rho^{\kappa}-\frac{l'(l'+1)}{\rho^{2}}\right]R_{nl}=0
\end{equation}
with the following \emph{ansatz}
\begin{equation}\label{ansatz4}
\varepsilon=\frac{\hbar^{2}E}{2mB^{2}}, \quad
\gamma^{2}=\frac{2mCr_{0}^{\kappa+2}}{\hbar^{2}},\quad
l'=-\frac{1}{2}+\sqrt{(l+\frac{1}{2})^{2}+\widetilde{A}}, \quad
\widetilde{A}=\frac{2mA}{\hbar^{2}}
\end{equation}
we can transform equation (\ref{schr}) to another Schr\"{o}dinger
equation form by changing the variable to $\rho =u^{2}$ and then
by inserting $R(u)=u^{\frac{1}{2}}\phi(u)$ into the transformed
equation. Therefore, we obtain the Schr\"{o}dinger equation as
follows

\begin{equation}\label{scrnum}
\frac{d^{2}\phi(u)}{du^{2}}+\left[4\varepsilon
u^{2}+4-4\gamma^{2}u^{2\kappa+2}
-\frac{\Lambda(\Lambda+1)}{u^{2}}\right]\phi(u)=0
\end{equation}
where $\Lambda=2l'+\frac{1}{2}$.

\subsection{$\kappa=1$ Case}
For $\kappa=1$, we obtain the following equation by using equation
(\ref{scrnum}):
\begin{equation}\label{diffk1}
\frac{d^{2}\phi(u)}{du^{2}}+\left[4\varepsilon u^{2}+4
-4\gamma^{2}u^{4} -\frac{\Lambda(\Lambda+1)}{u^{2}}\right]\phi(u)=0
\end{equation}
If we take the wave function in the following form
\begin{equation} \label{asymwWF}
\phi(u)=u^{\Lambda+1}exp(-\frac{\gamma\beta u^{4}}{2})f(u)
\end{equation}
where $\beta$ is an arbitrarily introduced constant to improve the
convergence speed of the method \cite{fernandez}. We also introduce
$\gamma$ in the asymptotic wave function since the parameter
$\gamma$ appears explicitly in the potential. It is known that when
changing the value of $\gamma$ in the potential, it affects the
eigenvalues and the shape of the wave function. Therefore, by
introducing $\gamma$ into equation (\ref{asymwWF}), we control the
change of the potential parameters and, as a result, obtain a better
convergence. If we insert this wave function into equation
(\ref{diffk1}), then we obtain the second-order homogeneous linear
differential equation as follows;
\begin{equation}\label{scraim1}
    \frac{d^2f(u)}{du^{2}}=\left[2(2\beta\gamma u^{3}-\frac{\Lambda+1}{u}) \right ]\frac{df(u)}{du}
    +\left[(4\beta\gamma\Lambda+10\beta\gamma -4\varepsilon)u^{2}-4\beta^{2}\gamma^{2}u^6-4+4\gamma^{2} u^{4}\right]f(u)
\end{equation}
which is now amenable to an AIM solution. By comparing this equation
with equation (\ref{diff}), we can write the $s_{0}(u)$ and
$\lambda_{0}(u)$ values as below
\begin{eqnarray}
s_{0}(u)&=&(4\beta\gamma\Lambda+10\beta\gamma -4\varepsilon)u^{2}-4\beta^{2}\gamma^{2}u^6-4+4\gamma^{2} u^{4} \\
\lambda_{0}(u)&=& 2(2\beta\gamma u^{3}-\frac{\Lambda+1}{u})
\end{eqnarray}
In order to obtain the energy eigenvalues from
equation~(\ref{scraim1}) by using equation~(\ref{iter}), we obtain
the $s_k(r)$ and $\lambda_k(r)$ in terms of $s_0(r)$ and
$\lambda_0(r)$. Then, by using the quantization condition of the
method given by equation (\ref{quantization}), we obtain the energy
eigenvalues. Therefore, we have to choose a suitable $u_{0}$ point
to solve the equation $\delta _{n}(u_{0},\varepsilon )=0 $
iteratively in order to find $\varepsilon $ values. In this study,
we obtain the $u_{0}$ from the maximum point of the asymptotic wave
function, which is the same as the root of $\lambda _{0}(u)=0$, thus
$u_{0}=\left(\frac{\Lambda+1}{2\beta\gamma }\right) ^{1/4}$. This
straightforward application of AIM gives us the energy eigenvalues
as the sequence appears to converge when the number of iterations
($k$) are $k \simeq$ 30 as shown in Table~\ref{Table1}.

In this table, we also present the convergence rate of AIM calculations.
The energy eigenvalues appear as the number of iterations are $k
\simeq$ 30. However, the speed of the convergence depends on the
arbitrarily introduced constant, $\beta$. We have investigated the
optimum values of $\beta$ that give the best convergence and have
kept the one that appears to yield the best convergence rate.
Therefore, for case $\kappa$=1, we have performed calculations for
the different values of $\beta$. It is seen from Table~\ref{Table1}
that the best convergence is obtained when the constant $\beta$
values are $\beta$=0.4, 0.5, and 0.6. For other values of $\beta$,
the convergence needs more iteration.

Having determined the value of $\beta$ in an empirical way, for case
$\kappa$=1, the energy eigenvalues by using AIM are shown for
different values of $n$ and $l$ for convergence constant
$\beta=0.5$, $\widetilde{A}=1$ and $\gamma=1$ in Table \ref{Table2}.
\subsection{$\kappa=2$ Case}
For $\kappa=2$, we obtain the following equation by using equation
(\ref{scrnum}):
\begin{equation}\label{diffk2}
\frac{d^{2}\phi(u)}{du^{2}}+\left[4\varepsilon u^{2}+4
-4\gamma^{2}u^{6} -\frac{\Lambda(\Lambda+1)}{u^{2}}\right]\phi(u)=0
\end{equation}
If we take the wave function in the following form
\begin{equation}
\phi(u)=u^{\Lambda+1}exp(-\frac{\gamma \beta u^{4}}{2})f(u)
\end{equation}
When we insert this wave function into equation (\ref{diffk2}), we
obtain the second-order homogeneous linear differential equation as
follows
\begin{equation}\label{scraim2}
    \frac{d^2f(u)}{du^{2}}=2\left[ (2\beta\gamma u^{3}-\frac{\Lambda+1}{u})      \right ]\frac{df(u)}{du}
 +\left[(4\beta\gamma\Lambda+10\beta\gamma-4\varepsilon)u^2-4\beta^{2}\gamma^{2}u^{6}-4+4\gamma^{2}u^{6}\right]f(u)
\end{equation}
which is now amenable to an AIM solution. By comparing this equation
with equation (\ref{diff}), we obtain $s_{0}(u)$ and
$\lambda_{0}(u)$ values as below
\begin{equation}
s_{0}= 4\beta\gamma
u^2\Lambda-4\beta^{2}\gamma^{2}u^{6}+10\beta\gamma u^2-4\varepsilon
u^2-4+4\gamma^{2}u^{6}
\end{equation}
\begin{equation}
\lambda_{0}=2(2\beta\gamma u^{3}-\frac{\Lambda+1}{u})
\end{equation}
Similar to case $\kappa=1$, in order to obtain the energy
eigenvalues from equation~(\ref{scraim2}) by using
equation~(\ref{iter}), we obtain the $s_k(r)$ and $\lambda_k(r)$ in
terms of $s_0(r)$ and $\lambda_0(r)$. Then, by using the
quantization condition of the method given by equation
(\ref{quantization}), we obtain the energy eigenvalues. For this
case, we choose $u=(\frac{\Lambda+1}{2\beta\gamma})^{\frac{1}{4}}$
and we obtain the energy eigenvalues after the $k=30$ iterations.
The AIM results are presented in Table \ref{Table3} for different
values of $n$ and $l$ quantum numbers for different values of
$\gamma$. In Table \ref{Table4} and \ref{Table5}, we compare our
results with previous works conducted for special forms of our
potential. Our results are in excellent agreement with moment method
\cite{moment}, which is the closest one to exact solution. They are
better than perturbation results.

\section{Conclusion}
\label{conclude}

In this paper, we have presented the exact and iterative solutions
of the radial Schr\" {o}dinger equation for a class of potential,
$V(r)=\frac{A}{r^{2}}-\frac{B}{r}+Cr^{\kappa}$, for various values
of $\kappa$ from -2 to 2 for any $n$ and $l$ quantum states by
applying the asymptotic iteration method. According to the value of
$\kappa$, this potential family includes Kratzer, Modified Kratzer,
Goldman-Krivchenkov or spiked harmonic oscillator, Coulomb plus
linear and Coulomb plus harmonic oscillator potentials.

These potentials have been extensively used to describe the bound
and the continuum states of the interactions systems and a great
number of papers have been published for the exact and numerical
solutions of these potentials (see Refs.
\cite{kratzer,fluge,swkb,k1,k2,k3,hakan,envolepe,Enrique,moment,shifted,Avron,Taut}
and references therein). We have examined these potentials in this
paper and have attempted to obtain exact or numerical solutions.

The global analysis of this potential family by using the asymptotic
iteration method has resulted in exact analytical solutions for the
values of $\kappa=0 ,-1$ and $-2$. In addition, closed-forms for
the energy eigenvalues as well as the corresponding eigenfunctions
are obtained, but for $\kappa=1$ and $2$, there are no analytical
solutions. Therefore, the energy eigenvalues are obtained
numerically.

The advantage of the asymptotic iteration method is that it gives
the eigenvalues directly by transforming the second-order
differential equation into a form of ${y}''$ =$ \lambda _0 (r){y}' +
s_0 (r)y$. The wave functions are easily constructed by iterating
the values of $s_0(r)$ and $\lambda_0(r)$. The asymptotic iteration
method results in exact analytical solutions if there is and
provides the closed-forms for the energy eigenvalues as well as the
corresponding eigenfunctions. Where there is no such a solution, the
energy eigenvalues are obtained by using an iterative approach
\cite{barakat,fernandez,orhan,orhanIJQC,hakanper,boztosun1}. As it
is presented, AIM puts no constraint on the potential parameter
values involved and it is easy to implement. The results are
sufficiently accurate for practical purposes. It is worth extending
this method to examine other interacting systems.

\section*{Acknowledgments}
This work is supported by the Scientific and Technical Research
Council of Turkey (T\"{U}B\.{I}TAK) under the project number
TBAG-2398 and Erciyes University (FBT-04-16).

\begin{table}[h]
\begin{center}
\begin{tabular}{ccccccccccccccc} \hline \hline
$k$ && $\beta=0.2$ && $\beta=0.4$&& $\beta=0.5$&& $\beta=0.6$&& $\beta=0.7$&& $\beta=0.9$&& $\beta=2.0$ \\\hline
 20 &&  2.36068441 && 2.36071158 && 2.36071387 && 2.36071387 && 2.36080271 && 2.36168203 && 2.46417111 \\
 30 &&  2.36071440 && 2.36071239 && 2.36071239 && 2.36071239 && 2.36071435 && 2.36076626 && 2.39133769 \\
 40 &&  2.36071234 &&     "      &&    "       &&     "      && 2.36071245 && 2.36071612 && 2.37021799 \\
 50 &&  2.36071282 &&     "      &&    "       &&     "      && 2.36071239 && 2.36071269 && 2.36372451 \\
 60 &&  2.36071253 &&     "      &&    "       &&     "      && 2.36071239 && 2.36071241 && 2.36168245 \\
 70 &&  2.36071268 &&     "      &&    "       &&     "      && 2.36071239 && 2.36071239 && 2.36102992 \\
 80 &&  2.36071240 &&     "      &&    "       &&     "      && 2.36071238 && 2.36071239 && 2.36081631 \\
 90 &&  2.36071238 &&     "      &&    "       &&     "      && 2.36071236 && 2.36071418 && 2.36076466 \\ \hline\hline
\end{tabular}
\end{center}
\caption{For $\kappa=1$ case, the energy eigenvalues ($\varepsilon$)
for the $n=0$ and $l=0$ states by means of several $\beta$ values.
We take $\widetilde{A}=1$ and $\gamma=1$.} \label{Table1}
\end{table}

\bigskip
\begin{table}[h]
\begin{center}
\begin{tabular}{ccccccccccc} \hline \hline
$k$ && $n=0, l=0$   && $n=1, l=0$ & $n=1, l=1$& $n=2, l=0$ &   & $n=2, l=1$& $n=2, l=2$&  \\\hline
 20 && 2.36071387   && 4.112474   & 4.7245997 & 5.57024    &   & 6.097373  & 6.739515  &  \\
 30 && 2.36071239   && 4.112295   & 4.7242771 & 5.55292    &   & 6.074553  & 6.706985  &  \\
 40 && 2.36071239   && 4.112291   & 4.7242690 & 5.55211    &   & 6.073379  & 6.704987  &  \\
 50 &&     "        && 4.112290   & 4.7242688 & 5.55208    &   & 6.073327  & 6.704887  &  \\
 60 &&     "        &&     "      &     "     & 5.55207    &   & 6.073324  & 6.704883  &  \\
 70 &&     "        &&     "      &     "     &     "      &   &     "     &     "     &  \\\hline\hline
\end{tabular}
\end{center}
\caption{For $\kappa=1$ case, the energy eigenvalues ($\varepsilon$)
for the several quantum states. We take $\widetilde{A}=1$,
$\gamma=1$ and convergence constant $\beta=0.5$.} \label{Table2}
\end{table}

\bigskip
\begin{table}[h]
\begin{center}
\begin{tabular}{rrrrrrrrrr} \hline \hline
$n$& & $l$&  & $\gamma=0.1$ &  & $\gamma=1$   &  &$\gamma=10$    &
\\\hline
 0 & &  0 &  & 0.1220043681 &  & 3.3582483393 &  & 39.6495973187 &   \\
   & &  1 &  & 0.3258602332 &  & 4.8963878137 &  & 53.8428825862 &   \\
   & &  2 &  & 0.5473302077 &  & 6.7964025443 &  & 72.0047051977 &   \\
 1 & &  0 &  & 0.5720226793 &  & 7.4731840675 &  & 79.9800946486 &   \\
   & &  1 &  & 0.7518135075 &  & 8.9639368105 &  & 94.0443338946 &   \\
   & &  2 &  & 0.9621813797 &  & 10.8379872414&  &112.1314195487 &   \\
 2 & &  0 &  & 0.9982832867 &  & 11.5427585197&  &120.1879082158 &   \\
   & &  1 &  & 1.1684333950 &  & 13.0102255287&  &134.1850120876 &   \\
   & &  2 &  & 1.372953445  &  & 14.8691668817&  &152.2273430525 &   \\ \hline\hline
\end{tabular}
\end{center}
\caption{For $\kappa=2$ case, the energy eigenvalues ($\varepsilon$)
for several quantum numbers. We take $\widetilde{A}=1$ and convergence constant $\beta=1$.}\label{Table3}
\end{table}

\begin{table}[h]
\begin{center}
\begin{tabular}{ccccccccccccccc} \hline \hline
$n$ && $l$ && $B$&& $C$&& Shifted $1/N$ $E$\cite{shifted} && Moment
$E$\cite{moment}&& AIM $E$\cite{barakat}&& Present Work $E$ \\
\hline 0   &&  0  &&  1 &&  1 &&  0.60025               &&  0.59377
&& 0.59365 &&    0.59377      \\ \hline\hline
\end{tabular}
\end{center}
\caption{Comparison of our results with the shifted $1/N$ expansion,
moment and perturbative asymptotic iteration methods for the energy
eigenvalues, where $A=0$, $B=1$, $C=1$, $\kappa=2$, m=1 and
$\hbar=1$.} \label{Table4}
\end{table}

\begin{table}[h]
\begin{center}
\begin{tabular}{ccccccccccccccc} \hline \hline
$C$   && $E$ \cite{moment}  && $E_{AIM}$ \\\hline
0.1         &&  -0.296088   &&-0.29608776      \\
0.5         &&  0.1796683   && 0.17966848      \\
1.0         &&  0.5937711   && 0.59377126      \\
2.0         &&  1.2237050   && 1.22370510      \\
5.0         &&  2.5617326   && 2.56173268      \\
10.0        &&  4.1501236   && 4.15012364      \\
20.0        &&  6.4799505   && 6.47995056      \\
50.0        &&  11.2654474  && 11.26544748      \\
100.0       &&  16.8052478  && 16.80524784      \\
1000.0      &&  59.3754689  && 59.37546904      \\
2000.0      &&  85.7348038  && 85.73480386      \\
5000.0      &&  138.5571975 && 138.55719764      \\\hline\hline
\end{tabular}
\end{center}
\caption{Comparison of our results with the ground state energy
eigenvalues of the hydrogen atom in the $V(r)=C r^{2}$ potential,
where $A=0$, $B=1$, $\kappa=2$, m=1 and $\hbar=1$.} \label{Table5}
\end{table}
\end{document}